\documentclass[aps,prb,nobibnotes,citeautoscript,twocolumn,superscriptaddress,footinbib,10pt]{revtex4-1} 
\usepackage[utf8]{inputenc}
\usepackage{amsmath,amssymb}
\usepackage{mathrsfs} 
\usepackage{color}
\usepackage{graphicx,color,colortbl}
\usepackage{rotating,array,tabularx,booktabs}
\usepackage{varwidth,xcolor}
\usepackage{placeins}
\usepackage{siunitx}
\usepackage{array,multirow}

\DeclareGraphicsExtensions{.pdf,.PDF,.png,.PNG}

\allowdisplaybreaks

\renewcommand{\vec}[1]{\boldsymbol{#1}}%

\newcommand{\dif}{\mathrm{d}}%
\newcommand{\Eins}{\mathbf{1}}%
\newcommand{\Nabla}{\vec{\nabla}}%
\newcommand{\abs}[1]{\lvert#1\rvert}%
\newcommand{\norm}[1]{\lVert#1\rVert}%
\newcommand{\ZT}[1]{\textquotedblleft#1\textquotedblright}%

\newcommand{\uc}{q}%
\newcommand{\uu}{\boldsymbol{\hat{u}}}%
\newcommand{\Peclet}{P\'eclet}%
\newcolumntype{Y}{>{\centering\arraybackslash}X}%

\begin{document}

\title{Pair-distribution function of active Brownian spheres in two spatial dimensions: simulation results and analytic representation}

\author{Julian Jeggle}
\affiliation{Institut f\"ur Theoretische Physik, Center for Soft Nanoscience, Westf\"alische Wilhelms-Universit\"at M\"unster, D-48149 M\"unster, Germany}

\author{Joakim Stenhammar}
\affiliation{Division of Physical Chemistry, Lund University, SE-221 00 Lund, Sweden}

\author{Raphael Wittkowski}
\email[Corresponding author: ]{raphael.wittkowski@uni-muenster.de}
\affiliation{Institut f\"ur Theoretische Physik, Center for Soft Nanoscience, Westf\"alische Wilhelms-Universit\"at M\"unster, D-48149 M\"unster, Germany}

\begin{abstract}
We investigate the full pair-distribution function of a homogeneous suspension of spherical active Brownian particles interacting by a Weeks-Chandler-Andersen potential in two spatial dimensions. The full pair-distribution function depends on three coordinates describing the relative positions and orientations of two particles, the P\'eclet number specifying the activity of the particles, and their mean packing density. This five-dimensional function is obtained from Brownian dynamics simulations. We discuss its structure taking into account all of its degrees of freedom. In addition, we present an approximate analytic expression for the product of the full pair-distribution function and the interparticle force. We find that the analytic expression, which is typically needed when deriving analytic models for the collective dynamics of active Brownian particles, is in good agreement with the simulation results. The results of this work can thus be expected to be helpful for the further theoretical investigation of active Brownian particles as well as nonequilibrium statistical physics in general.  
\end{abstract}

\maketitle

\section{Introduction}
In the last two decades, so-called active matter has been of increasing interest in several fields of research including physics\cite{VicsekCBJCS1995, GrossmanABJ2008, CatesT2013}, chemistry\cite{PaxtonEtAl2004, WangDASM2015, WuWHLSH2013}, and biology\cite{LiuEtAl2011, RiedelKH2005, Goldstein2015}. The notion of activity refers to the fact that, unlike traditional systems of particles, the constituents of active matter are subject to self-propulsion. Examples of active systems can be found in the form of flocks and swarms of animals\cite{CavagnaG2014, CouzinKFL2005}, swimming microorganisms\cite{CzirkBJCV1996, DombrowskiCCGK2004, TailleurC2008, BarryB2010}, active cytoskeletal filaments\cite{NedelecSML1997, CamaletJP1999, SurreyNLK2001, KruseJJPS2004, BackoucheHGBG2006, SchallerWSFB2010}, and suspensions of artificial particles propelled by a variety of mechanisms\cite{BricardCDDB2013, PalacciSSPC2013, ButtinoniBKLBS2013, YanHZXLG2016, NiuPS2017}. The active motion gives rise to a number of intriguing many-particle phenomena that are enabled by the intrinsically nonequilibrium nature of active systems. Detailed reviews on these topics can be found in Refs.\ \onlinecite{Ramaswamy2010, RomanczukBELSG2012, MarchettiJRLPRS2013, BialkeSL2015, Menzel2015, ElgetiWG2015, BechingerdLLRVV2016}. One particularly noteworthy phenomenon is motility-induced phase separation, where a suspension of active particles spontaneously separates into a dense \ZT{cluster} phase and a dilute \ZT{gas} phase despite equal and purely repulsive interactions between the particles\cite{RednerHB2013, BialkeLS2013, StenhammarMAC2014, CatesT2015, FarageKB2015}.

When studying many-particle systems, the pair-distribution function, which describes the probability for finding two particles with a particular configuration simultaneously, is of great importance, since it is fundamentally related to the macroscopic properties of the system and, e.g., often needed when deriving field theories for such systems.\cite{HansenMD2009} For systems of passive (i.e., nonmotile) particles, the pair-distribution function has already been studied in detail by experiments\cite{vanBlaaderenW1995, CarbajalTinocoCRAL1996, Hughes2010, IacovellaRGS2010, ThorneyworkRAD2014} and computer simulations\cite{IacovellaRGS2010, AllenT2017}. Furthermore, analytic approaches like integral equation theory exist that allows for the calculation of approximate expressions for the pair-distribution function \cite{MoritaH1960, Percus1964, Stell1964, HansenMD2009}.
In the case of active systems, however, our understanding of the pair-distribution function is much less developed, since the pair-distribution function is then much more complicated than for passive particles which typically lack orientational degrees of freedom. Thus, most results from experiments, simulations, and theory address only approximate, reduced versions of this function\cite{BialkeLS2013, FarageKB2015, MarconiPM2016, ReinS2016, HaertelRS2018, PessotLM2018, SchwarzendahlM2019}, where the orientational degrees of freedom are often integrated out, instead of the \ZT{full} pair-distribution function including both the spatial arrangement of the particles and their orientations.
Nevertheless, the orientational degrees of freedom contain significant information about the particle dynamics. This can be seen, e.g., by a relatively large mismatch between the predicted spinodal corresponding to the onset of motility-induced phase separation, which is based on a reduced pair-distribution function, and simulation results indicating the true spinodal in Ref.\ \onlinecite{WittkowskiSC2017}. In fact, in the more rigorous derivation of a field theory in Ref.\ \onlinecite{BickmannW2019}, which is based on the full pair-distribution function, terms depending on the orientational degrees of freedom appear naturally and the predicted spinodal is in very good agreement with simulation results.

Examples of previous investigations of the pair-distribution function, where the same model of active Brownian particles as used in our present article is employed and numerical results for a reduced form of the pair-distribution function are presented, can be found in Refs.\ \onlinecite{BialkeLS2013,SchwarzendahlM2019, ReinS2016}. Similar results also exist for different models such as binary mixtures of particles with different propulsion mechanisms or mixtures of active and passive particles.\cite{PessotLM2018, WittkowskiSC2017} Particularly noteworthy is Ref.\ \onlinecite{HaertelRS2018}, where a theory for systems of active hard disks was developed and then compared to simulations of nearly hard spherical particles that are used also in the present article. While some results for the full pair-distribution function are shown, fitted analytic expressions are given only for a reduced form of the pair-distribution function and for a limited subset of the system parameters. 
Nevertheless, theoretical treatments of systems of active particles require knowledge of the full pair-distribution function. This applies especially to the derivation of field theories for active matter, where usually a product of the pair-distribution function and the interparticle force occurs\cite{BialkeLS2013, SpeckMBL2015, SteffenoniFK2017, WittkowskiSC2017, ReinkenKBH2018, BickmannW2019}. The relatively low number of field theories for active matter and the strong approximations involved in the theories developed in Refs.\ \onlinecite{BialkeLS2013, SpeckMBL2015, SteffenoniFK2017, WittkowskiSC2017, ReinkenKBH2018} can be attributed to an insufficient knowledge about the pair-distribution function for the active-particle systems.

In this article, we therefore provide further insights into the largely unknown structure of pair-distribution functions for active-matter systems. For this purpose, a homogeneous system of spherical active Brownian particles (ABPs)\cite{RomanczukBELSG2012,WensinkLMHWZKM2013,CatesT2015,BechingerdLLRVV2016,Speck2016,MalloryVC2018} that move in a plane and interact via the Weeks-Chandler-Andersen potential is addressed. This is an important standard system considered in many previous studies\cite{RednerHB2013, BialkeLS2013, ButtinoniBKLBS2013, StenhammarMAC2014, SpeckBML2014}. Based on Brownian dynamics simulations, we simulate the time evolution of the system and calculate the full pair-distribution function for homogeneous stationary states. This function depends on three coordinates, a radial distance and two angles, as well as the P\'eclet number Pe specifying the activity of the particles, and their mean packing density $\Phi_0$. We present and discuss the structure of this five-dimensional pair-distribution function. The results reveal a complex structure with a strong dependence on all arguments, showing that the full function needs to be carefully taken into account in theoretical modeling. In addition, we give an analytic approximation for the product of the pair-distribution function and the interparticle force, which is frequently needed in theoretical modeling. 

This article is organized as follows: In Sec.\ \ref{sec:methods}, we give an overview of our simulations and the calculation of the pair-distribution function. The results of our simulations and the analytic approximation are presented in Sec.\ \ref{sec:results}. Finally, we state our conclusions in Sec.\ \ref{sec:conclusions}.

\section{\label{sec:methods}Methods}
We describe the motion of $N$ active Brownian spheres using the overdamped Langevin equations \cite{BialkeLS2013,RednerHB2013,StenhammarMAC2014}
\begin{gather} 
\dot{\vec{r}}_{i} = \frac{D_T}{k_{\mathrm{B}}T} \Big( F_A \uu (\varphi_i) - \sum^{N}_{\begin{subarray}{c}j=1\\j \neq i\end{subarray}} \Nabla_{\vec{r}_i} U_{2}(\norm{\vec{r}_i-\vec{r}_j}) \Big) + {\vec{\xi}}_{T,i} , \raisetag{2ex}\\
\dot{\varphi}_i = \xi_{R,i}  
\end{gather}
with the position $\vec{r}_i(t)$ and orientation $\varphi_i(t)$ of particle $i$ at time $t$, translational diffusion coefficient $D_T$, Boltzmann constant $k_{\mathrm{B}}$, absolute temperature $T$, active force magnitude $F_A$, orientational unit vector $\uu(\varphi)=(\cos(\varphi),\sin(\varphi))^{\mathrm{T}}$, pair-interaction potential $U_{2}(r)$, and zero-mean Gaussian white noise terms $\vec{\xi}_{T,i}(t)$ and $\xi_{R,i}(t)$. The latter terms are normalized such that
\begin{gather}
\langle \vec{\xi}_{T,i}(t) \otimes \vec{\xi}_{T,j}(t') \rangle = 2D_T \Eins \delta_{ij} \delta(t-t'), \\
\langle \xi_{R,i}(t) \xi_{R,j}(t') \rangle = 2D_R \delta_{ij} \delta(t-t'),
\label{eq:langevin_eom}%
\end{gather}
where $\otimes$ denotes the dyadic product, $\Eins$ is the identity matrix, and $D_R$ is the rotational diffusion coefficient, which, for spherical particles, can be related to the translational diffusion coefficient given by the Stokes-Einstein relation $D_T=k_BT/(3\pi\eta\sigma)$ via the Stokes-Einstein-Debye relation $D_R = 3D_T/\sigma^2=k_BT/(\pi\eta\sigma^3)$, where $\sigma$ is the particle diameter and $\eta$ is the dynamic viscosity of the solvent surrounding the particles. A measure of the directional active motion compared to the random Brownian motion is given by the dimensionless \Peclet{} number $\mathrm{Pe}=\sigma F_A/(k_{\mathrm{B}}T)$. In our simulations, we studied systems with varying Pe and packing density $\Phi_0=\rho\pi\sigma^2/4$, where $\rho$ is the particle number density in the system. To ensure an equal effective particle radius across all simulations, Pe was controlled by a change in $T$ instead of $F_A$ \cite{StenhammarMAC2014}. Since the temperature diverges when approaching small Pe, our analysis covers the range $\textrm{Pe} \in [10,250]$. The packing density has both an upper bound at $\pi/(2\sqrt{3}) \approx 0.91$ due to reaching a dense circle packing with little to no room for motility as well as a lower bound due to the number of particles approaching zero. Therefore, we used the range $\Phi_0 \in [0.01,0.9]$ for the packing density. The interaction potential is described by the Weeks-Chandler-Andersen potential
\begin{equation}
U_{2}(r)=
\begin{cases}
4\varepsilon \Big( \big( \frac{\sigma}{r} \big)^{12} - \left( \frac{\sigma}{r} \right)^{6} \Big) + \varepsilon, & \mbox{if } r \leq 2^{1/6} \sigma, \\
0, & \mbox{else}
\end{cases}
\label{eq:wca}%
\end{equation}
with the scaling factor $\varepsilon$.

\begin{figure}[htb]
\centering
\includegraphics[]{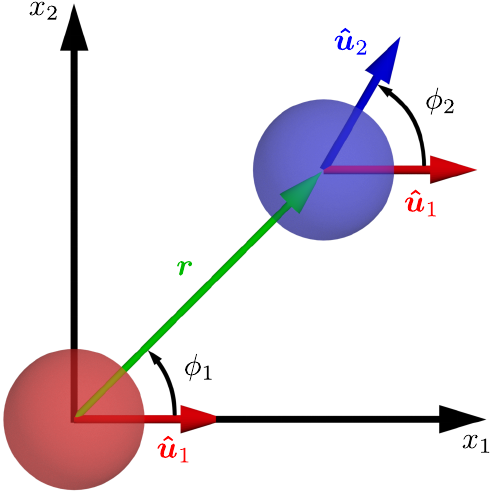}%
\caption{\label{fig:corrfunc_coords} Local coordinate system for the parameterization of the pair-distribution function $g(r,\phi_1,\phi_2)$. The vector $\vec{r}=\vec{r}_2-\vec{r}_1$ with length $r=\norm{\vec{r}}$ points from particle $1$ to particle $2$ and the unit vectors $\uu_1=\uu(\varphi_1)$ and $\uu_2=\uu(\varphi_2)$ denote the orientations of the particles.}
\end{figure}

The full pair-distribution function $g(\vec{r}_1,\varphi_1,\vec{r}_2,\varphi_2,t)$ depends on seven coordinates including time. If the system is in a stationary state, the time dependence of the pair-distribution function can be neglected. In the case of a homogeneous system, its translational symmetry can be used to reduce the dependence of $g$ on $\vec{r}_1$ and $\vec{r}_2$ to one on $\vec{r}_2-\vec{r}_1$. Similarly, the isotropy of the homogeneous system can be used to replace the dependence on $\varphi_1$ and $\varphi_2$ by one on $\varphi_2-\varphi_1$, thus reducing the number of coordinates of $g$ to three. Through the introduction of the local coordinate system shown in Fig.\ \ref{fig:corrfunc_coords}, where the vector $\uu_1=\uu(\varphi_1)$ that denotes the orientation of a particle 1 is parallel to the first axis $x_1$ of the particle-fixed coordinate system, the pair-distribution function can be reparameterized as $g(r,\phi_1,\phi_2)$. The arguments in this reparameterization are the center-to-center distance $r=\norm{\vec{r}_2-\vec{r}_1}$ between particles 1 and 2, the angle $\phi_1=\varphi_r-\varphi_1$ between the vectors $\uu_1$ and $\vec{r}=\vec{r}_2-\vec{r}_1$, where the angle $\varphi_r$ is defined by the equation $\uu(\varphi_r)=\vec{r}/\norm{\vec{r}}$, and the angle $\phi_2=\varphi_2-\varphi_1$ between the orientations $\uu_1$ and $\uu_2=\uu(\varphi_2)$ of particles 1 and 2. With respect to $\phi_1$ and $\phi_2$, $g(r,\phi_1,\phi_2)$ is periodic with period $2\pi$. The reparameterized pair-distribution function furthermore has the symmetry
\begin{equation}
g(r, \phi_1, \phi_2) = g(r, -\phi_1, -\phi_2)
\label{eq:corrfunc_sym}%
\end{equation}
as can be seen from Fig.\ \ref{fig:corrfunc_coords}. This symmetry allows us to mirror our numerical data about $\phi_1=0$ and $\phi_2=0$ to reduce numerical noise.
Of particular relevance for the development of field theories that describe the collective dynamics of systems of ABPs is the function $-g(r,\phi_1,\phi_2)U'_2(r)$, where $-U'_2(r)=-\dif U_2(r)/\dif r$ is the interparticle force \cite{BialkeLS2013,StenhammarWMC2015,WittkowskiSC2017,BickmannW2019}. This product function can be seen as a ``pair-interaction-force distribution" and plays an essential role in the structure and dynamics of any many-body system with pairwise interactions. \cite{WittkowskiSC2017, BickmannW2019, BialkeLS2013, ReinkenKBH2018, SpeckMBL2015, WittkowskiL2011,Bickmann2019a,BickmannBJW} In general, the pair-distribution function $g(r,\phi_1,\phi_2)$ has a complex structure that is difficult to express analytically. It is thus advantageous to search for an approximation of the product function $-g(r,\phi_1,\phi_2)U'_2(r)$ instead, where a much smaller range of values for $r$ has to be considered: While the interparticle force $-U'_2(r)$ is zero for $r\geq r_{\mathrm{max}}=2^{1/6}\sigma$, the pair-distribution function $g(r,\phi_1,\phi_2)$ is zero for small $r$ due to the strong repulsion of particles at short distances. In our simulations, the smallest value of $r$ where $g$ is nonzero was $r_{\mathrm{min}}\approx 0.78 \sigma$. Thus, only a small support remains where the product function is nonzero, greatly simplifying its description.

To objectively judge the onset of clustering, the characteristic length scale $L_C$ was calculated, defined as\cite{StenhammarMAC2014}
\begin{equation}
L_C=2\pi\frac{\int_{2\pi/\ell}^{k_{\mathrm{cut}}}S(k) \,\dif k}{\int_{2\pi/\ell}^{k_{\mathrm{cut}}}kS(k) \,\dif k}\;,
\label{eq:charlen}%
\end{equation}
where $\ell$ is the edge length of the quadratic simulation domain, $k_{\mathrm{cut}}$ is the upper cutoff for the modulus $k=\norm{\vec{k}}$ of the wave vector $\vec{k}$, and $S(k)$ is the structure factor. The cutoff was set to $k_\mathrm{cut}=\pi$, which approximately coincides with the first minimum of $S(k)$. Only the vectors $\vec{k}$ conforming to the periodicity of the simulation domain have to be considered, which simplifies the integral over $k$ to a discrete sum. An in-depth analysis of the structure factor for active hard disks in two spatial dimensions can be found in Ref.\ \onlinecite{deMacedoBiniossekLVS2018}.

Throughout the article we use Lennard-Jones units, where $\varepsilon$, $\sigma$, and the Lennard-Jones time $\tau_{\mathrm{LJ}}$ are chosen as units of energy, length, and time, respectively. Furthermore, we set the active force to $F_A=24\varepsilon/\sigma$ and the particle mobility to $D_T/(k_{\mathrm{B}}T)=\sigma^{2}/(\tau_{\mathrm{LJ}}\varepsilon)$, which, since we tune Pe via $T$, remains constant. The numerical results were obtained using a modified version of the molecular dynamics simulation package LAMMPS\cite{Plimpton1995}. A system size of $\ell=256\sigma$ and simulation times of $2500\tau_{\mathrm{LJ}}$ with a time step $\Delta t = 5 \cdot 10^{-5} \tau_{\mathrm{LJ}}$ were used. The pair-distribution function was recorded with a resolution of 180 data points for the angles $\phi_1$ and $\phi_2$ each and 2000 data points for $r\in[0,10\sigma]$.

\section{\label{sec:results}Results and discussion}
\begin{figure}[htb]
\centering
\includegraphics[]{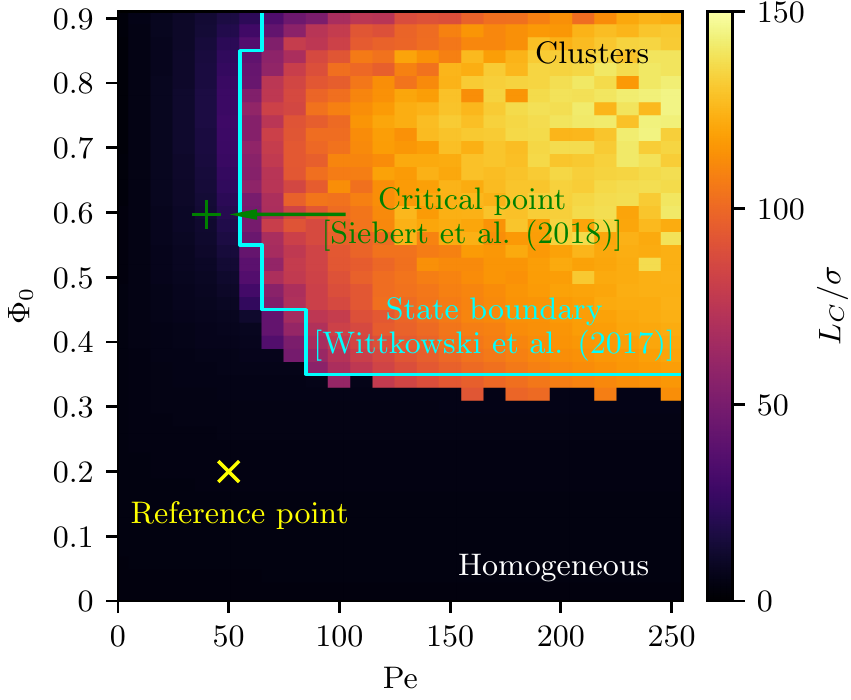}%
\caption{\label{fig:phase_diagram}Characteristic length $L_C$ as a function of \Peclet{} number $\mathrm{Pe}$ and packing density $\Phi_0$. The regions where the system stays homogeneous and where clusters form can easily be distinguished. The solid line shows previous simulation results for the parameters separating the region of spontaneous clustering, reproduced from Ref.\ \onlinecite{WittkowskiSC2017}, the green plus an estimate for the critical point at $\mathrm{Pe}=40$ and $\Phi_0=0.60$ according to Ref.\ \onlinecite{SiebertDSBSV2018}, and the yellow cross the reference point given by the parameter combination with $\mathrm{Pe}=50$ and $\Phi_0=0.2$ to which the later figures in this article correspond. Data points at $\Phi_0=0$ and $\mathrm{Pe}=0$ are extrapolated from simulations down to $\Phi_0=0.01$ and $\mathrm{Pe}=10$, respectively. A file containing the data for $L_C$ shown in this state diagram as well as two movies of the time evolution of the system corresponding to a point in each region of the state diagram are provided as Supplementary Material.}
\end{figure}

Since the parameterization $g(r,\phi_1,\phi_2)$ of the pair-distribution function requires the system to be in a homogeneous state, we first consider the system's state diagram. For this purpose, we performed simulations for a grid of parameter combinations with spacings $\Delta\Phi_0 = 0.02$ and $\Delta\textrm{Pe} = 10$. During the simulations, the characteristic length $L_C$ defined by Eq.\ \eqref{eq:charlen} was calculated and averaged over time, neglecting early times where the steady state of the system had not yet been reached. 
The resulting state diagram is presented in Fig.\ \ref{fig:phase_diagram}. 
It is clearly in line with the state diagram shown in Ref.\ \onlinecite{WittkowskiSC2017}, where the resolution was lower and the different states were distinguished by visual inspection. In our state diagram, one can see a very sharp transition from a homogeneous region to a cluster region at $\Phi_0=0.32$ for $\mathrm{Pe}\geq 170$. For lower \Peclet{} numbers, the change in characteristic length is much more gradual, which is a consequence of the fluctuations around the critical point that lies approximately at $\mathrm{Pe}=40$ and $\Phi_0=0.60$ \cite{SiebertDSBSV2018}. To avoid a considerable influence of these fluctuations as well as the occurrence of inhomogeneous steady states, we excluded the parameter combinations with $\mathrm{Pe}>30$ and $\Phi_0>0.3$ from our analysis of the pair-distribution function that we describe in the following.

\subsection{\label{sec:results:corr_func}Pair-distribution function}
\begin{figure*}[htb]
\centering
\includegraphics[]{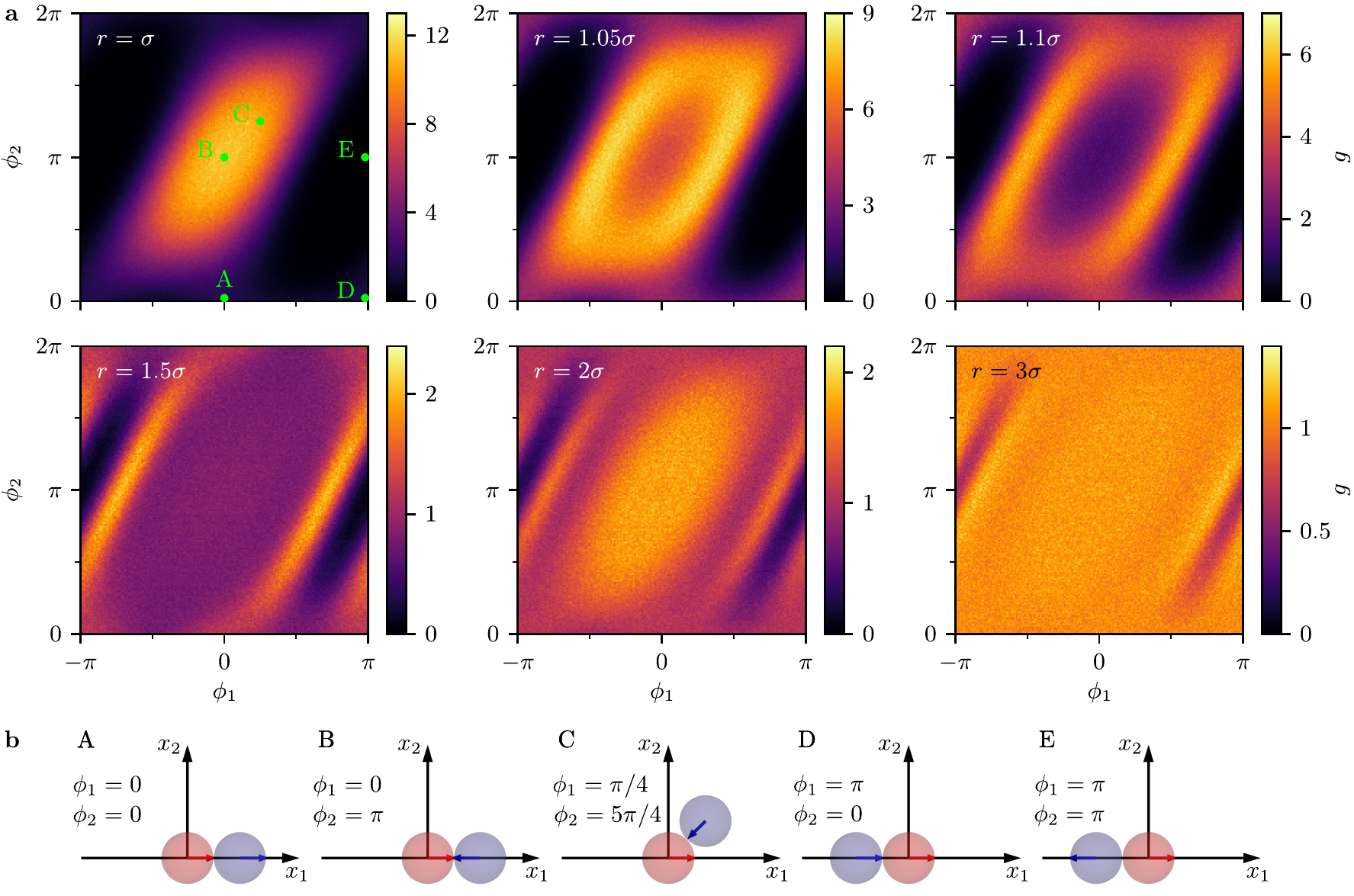}
\caption{\label{fig:corr_slices}(a) Pair-distribution function $g(r,\phi_1,\phi_2)$ for selected distances $r$ and the reference parameters $\mathrm{Pe}=50$ and $\Phi_0=0.2$ (see Fig.\ \ref{fig:phase_diagram}) as well as (b) sketches of the particle configurations marked in the top left plot. A movie showing $g(r,\phi_1,\phi_2)$ as a function of $\phi_1$ and $\phi_2$, where $r$ increases over time, is included in the Supplementary Material.}
\end{figure*}

In Fig.\ \ref{fig:corr_slices}, the pair-distribution function $g(r,\phi_1,\phi_2)$ is shown for a few values of $r$. 
The general structure and especially the extrema in correlation can be understood from the relative distances and orientations of two particles. A maximum in correlation is to be expected where two particles remain for longer times than average, while minima correspond to configurations that are either hard to reach or particularly short-lived. For example, at $r\approx\sigma$ a maximum exists for $\phi_1=0$ and $\phi_2=\pi$, which corresponds to the configuration of two particles with opposite orientations that inhibit each other's motion and thus remain relatively long in this configuration. In contrast, the configuration with $\phi_1=\phi_2=\pi$, where the secondary particle is behind the primary particle with an opposite orientation, represents a minimum in $g$, since the secondary particle would have to move through the primary particle to achieve such a configuration. The skewed structure of the pair-distribution function can be explained by the higher stability of configurations where the secondary particle is oriented in parallel to the vector pointing from the first to the second particle (see, e.g., in Fig.\ \ref{fig:corr_slices} the configuration with $\phi_1=\pi/4$ and $\phi_2=5\pi/4$) compared to configurations where these vectors are perpendicular to each other. 

For larger distances, the singular maximum turns first into a ring-like structure and later into two bands of increased correlation that move outwards from $\phi_1=0$. At distances equal to multiples of $\sigma$, this pattern repeats, albeit with a lower intensity. This periodicity is a result of shell-like arrangements of particles also observed in the radial distribution function of passive particles. The minimum at $\phi_1=\phi_2=\pi$ lies at the center of a spot of minimal correlation that becomes squeezed when $r$ increases. Compared to the maximum at $\phi_1=0$ and $\phi_2=\pi$, the minimum fades slower when $r$ increases as can be seen for $r=3\sigma$ in Fig.\ \ref{fig:corr_slices}.

\subsection{\label{sec:results:approx}Analytic approximation for the function $\boldsymbol{-gU_{2}'}$}
The product function $f(r,\phi_1,\phi_2)=-g(r,\phi_1,\phi_2)U_{2}'(r)$ depends on the coordinates $r$, $\phi_1$, and $\phi_2$ as well as the \Peclet{} number $\mathrm{Pe}$ and the packing density $\Phi_0$. To represent this five-dimensional function by an approximate analytic expression, we first perform an expansion into a Fourier series with respect to the angles $\phi_1$ and $\phi_2$. This Fourier series is truncated at second order, since we found this order to be sufficient to approximate the structure of the pair-distribution function $g(r,\phi_1,\phi_2)$ (see Fig.\ \ref{fig:corr_slices}) with reasonable accuracy. 
\begin{figure*}[htb]
\centering
\includegraphics[]{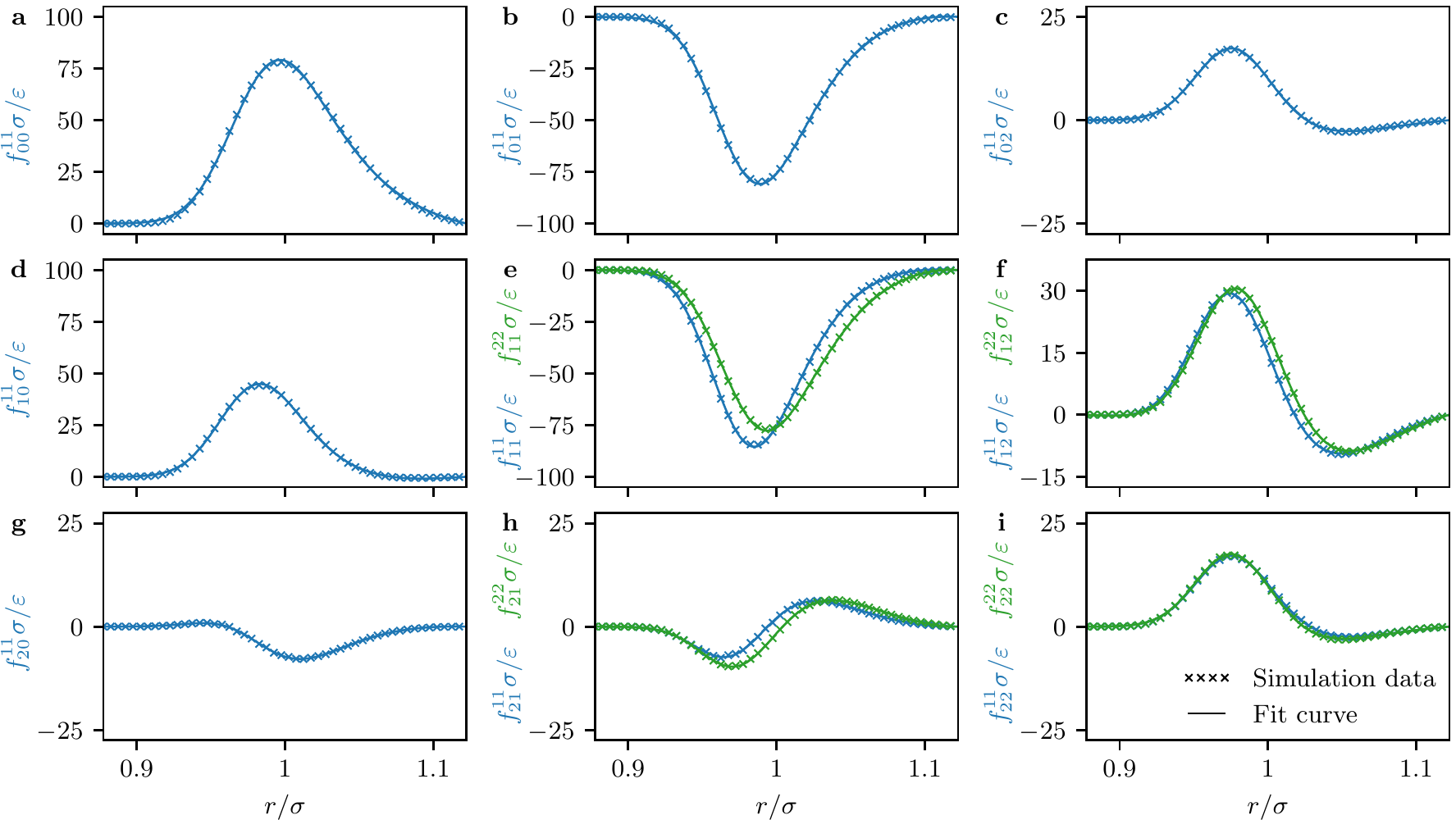}%
\caption{\label{fig:fourier_coeffs}Simulation data and corresponding fitted curves described by Eq.\ \eqref{eq:fit} for the Fourier coefficients $f_{I}(r)$ of the function $f(r,\phi_1,\phi_2)$ for the reference parameters $\mathrm{Pe}=50$ and $\Phi_0=0.2$. See the Appendix for a detailed list assigning the fit functions to the individual Fourier coefficients.}
\end{figure*}
Our approximation of $f(r,\phi_1,\phi_2)$ is therefore given by 
\begin{equation}
f(r,\phi_{1},\phi_{2})\approx\sum_{k,l=0}^{2}\sum_{i,j=1}^{2}f_{kl}^{ij}(r)u_{i}(k\phi_{1})u_{j}(l\phi_{2})
\label{eq:Fourier_approximation}%
\end{equation}
with the Fourier coefficients
\begin{equation}
\begin{split}
f_{kl}^{ij}(r)=\,&\frac{1}{(\delta_{k0}+1)(\delta_{l0}+1)}\int_{-\pi}^{\pi} \hspace{-2.2ex}\dif\phi_{1}\int_{-\pi}^{\pi} \hspace{-2.28ex}\dif\phi_{2} \\
&f(r,\phi_{1},\phi_{2})u_{i}(k\phi_{1})u_{j}(l\phi_{2})
\end{split}
\end{equation}
and the vector elements $u_{i}(\varphi)=(\uu(\varphi))_{i}$.
As a consequence of the symmetry property \eqref{eq:corrfunc_sym} of $g(r,\phi_1,\phi_2)$, some of the Fourier coefficients vanish: 
\begin{equation}
f^{12}_{kl}(r)=f^{21}_{kl}(r)=0 \qquad \forall k,l .
\end{equation}
We therefore simplify our notation by introducing the short form $f_{I}(r)\equiv f_{kl}^{ii}(r)$ with a multiindex $I=(i,k,l)$.
The five-dimensional function $f(r,\phi_{1},\phi_{2})$ is thus represented by 13 Fourier coefficients $f_{I}(r)$ that depend on $r$, $\mathrm{Pe}$, and $\Phi_0$.

To achieve the wanted analytic representation of the product function, we need to replace the discrete tabulation of Fourier coefficients obtained from our simulations by continuous functions in $r$, $\mathrm{Pe}$, and $\Phi_0$. For this purpose, we searched empirically for suitable functions reproducing the general shape and features of the curves for the Fourier coefficients shown in Fig.\ \ref{fig:fourier_coeffs}. We observed that the bell-shaped Fourier coefficients $f_{I}(r)$ can be fitted reasonably well with the help of the exponentially modified Gaussian distribution 
\begin{align}
\begin{split}
\mathrm{EMG}(r;\mu,\varsigma,\lambda) &= \frac{\lambda}{2}\exp\!\bigg( \frac{\lambda}{2}\big(\lambda\varsigma^{2}-2(r-\mu)\big)\!\bigg)\\
&\quad\;\, \mathrm{erfc}\bigg( \frac{\lambda\varsigma^{2}-(r-\mu)}{\sqrt{2}\varsigma} \bigg) ,
\end{split}
\end{align}
where $\mu$ is the mean of the distribution, $\varsigma$ is its standard deviation, $\lambda$ describes a skew in the distribution, and $\mathrm{erfc}(x)$ denotes the complementary error function. Since the function $f(r,\phi_1,\phi_2)$ must be equal to zero for $r \geq r_{\mathrm{max}}$ and we did not observe more than two zero-crossings for $r < r_{\mathrm{max}}$ in the numerical data, one can use the following set of functions to fit all Fourier coefficients $f_{I}(r)$:
\begin{equation}
\begin{split}
f_{I}(r) & \approx f_{0,I}\text{EMG}(r;\mu_{I},\varsigma_{I},\lambda_{I})\Theta(r_{\text{max}}-r)\\
&\quad\,\:\! (r-r_{\text{max}})\prod_{m=1}^{M_{I}}(r-r_{m,I}).
\end{split}\label{eq:fit}%
\end{equation}
These fit functions include a scaling factor $f_{0,I}$, parameters $\mu_{I}$, $\varsigma_{I}$, and $\lambda_{I}$, the Heaviside step function $\Theta(x)$ for cutting off $f_I(r)$ at $r=r_{\text{max}}=2^{1/6}\sigma$, and a total of $M_I$ additional roots $r_{m,I}$ with
\begin{equation}
(M_{I}) \equiv (M_{ikl})_{k,l=1,2,3} 
= \begin{pmatrix}
0 & 1 & 1 \\
1 & 1 & 2 \\
2 & 2 & 1 \\
\end{pmatrix}
\end{equation}
to capture the observed zero-crossings in the Fourier coefficients.
The number of zero-crossings of some coefficients $f_{I}(r)$ varies with $\mathrm{Pe}$ and $\Phi_0$. In such cases, the fit function was chosen according to the maximum number of zero-crossings observed over all $\mathrm{Pe}$ and $\Phi_0$. When the numerical data do not support this number of zeros, some fit parameters $r_{m,I}$ move to values either much smaller than $r_{\mathrm{min}}$ or larger than $r_{\mathrm{max}}$. Example fits of the Fourier coefficients are shown in Fig.\ \ref{fig:fourier_coeffs}. An interesting observation that can be made is the strong structural similarity of the coefficients $f^{11}_{kl}(r)$ and $f^{22}_{kl}(r)$.

\begin{figure*}[htb]
\centering
\includegraphics[width=\linewidth]{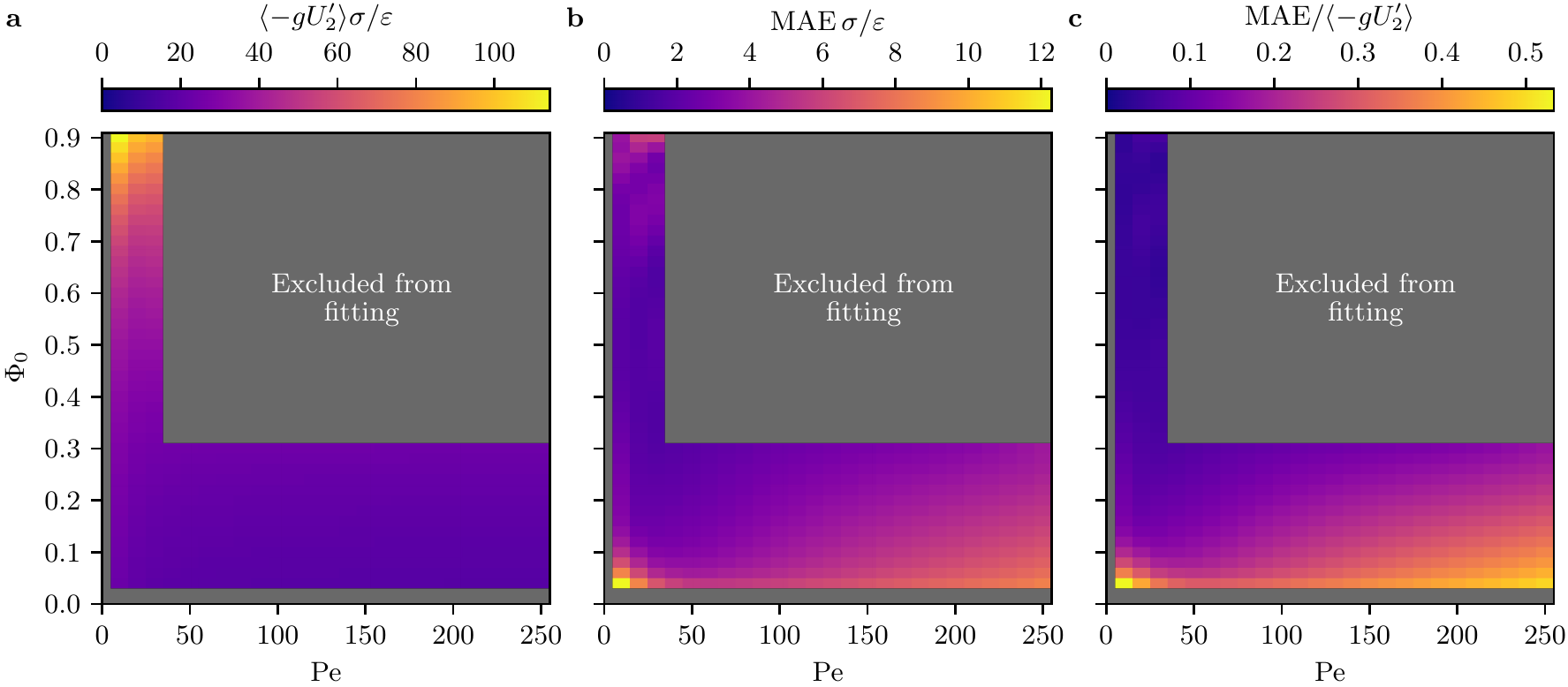}%
\caption{\label{fig:fit_error}(a) Mean absolute value $\langle -gU_{2}'\rangle$ of the function $-g(r,\phi_1,\phi_2)U_{2}'(r)$ determined by the simulations, (b) mean absolute error (MAE) of the analytic approximation for $-g(r,\phi_1,\phi_2)U_{2}'(r)$ compared to the corresponding simulation results, and (c) relative error $\mathrm{MAE}/\langle -gU_{2}'\rangle$.}
\end{figure*}

The two approximation steps performed so far have reduced the five-dimensional function $f(r,\phi_1,\phi_2)$ to an analytic expression including the functions $f_{0,I}$, $\mu_{I}$, $\varsigma_{I}$, $\lambda_{I}$, and $r_{m,I}$ that depend on $\mathrm{Pe}$ and $\Phi_0$. 
When determining $f_{0,I}$, $\mu_{I}$, $\varsigma_{I}$, $\lambda_{I}$, and $r_{m,I}$ as fit parameters in Eq.\ \eqref{eq:fit} for various values of $\mathrm{Pe}$ and $\Phi_0$, it is important to make sure that they vary continuously with $\mathrm{Pe}$ and $\Phi_0$. This is necessary in order to interpolate continuously between the sample points for different values of $\mathrm{Pe}$ and $\Phi_0$ and to obtain eventually a rigorous analytic expression for the product function.
Then, each of the functions $f_{0,I}$, $\mu_{I}$, $\varsigma_{I}$, $\lambda_{I}$, and $r_{m,I}$ of $\mathrm{Pe}$ and $\Phi_0$ can be well approximated by one of the expressions
\begin{align}
h_{m}(\mathrm{Pe},\Phi_{0})&=\sum_{i=-2}^{2}\sum_{j=0}^{m}\uc_{i,j} \mathrm{Pe}^{\frac{i}{2}} \Phi_{0}^{j} ,\label{eq:hm}\\
h_{m,n}(\mathrm{Pe},\Phi_{0})&=h_{m}(\mathrm{Pe},\Phi_{0})+v \mathrm{Pe}^{-n} e^{w\Phi_0} \label{eq:hmn}%
\end{align}
with the fit coefficients $\uc_{i,j}$, $v$, and $w$. Once more, this ansatz is empirically motivated and focuses on reproducing the general shape of the parameters as functions of $\mathrm{Pe}$ and $\Phi_0$. Negative powers of $\mathrm{Pe}$ have to be allowed in the expressions \eqref{eq:hm} and \eqref{eq:hmn} to reproduce the divergence of many parameters in the limit $\mathrm{Pe}\to 0$. In particular, we used $h_2(\mathrm{Pe},\Phi_{0})$, which contains 15 coefficients, $h_3(\mathrm{Pe},\Phi_{0})$ with 20 coefficients, as well as $h_{2,0}(\mathrm{Pe},\Phi_{0})$ and $h_{2,1}(\mathrm{Pe},\Phi_{0})$ with 17 coefficients each. The results of this fitting procedure are shown in the Appendix.

To determine the quality of our analytic approximation $f_{\mathrm{app}}(r,\phi_1,\phi_2;\mathrm{Pe},\Phi_0)$ of the product function $f(r,\phi_1,\phi_2;\mathrm{Pe},\Phi_0)=-g(r,\phi_1,\phi_2;\mathrm{Pe},\Phi_0)U_{2}'(r)$ determined by the simulations, we calculated the mean absolute error of $f_{\mathrm{app}}$ compared to $f$: 
\begin{equation}
\mathrm{MAE}(\mathrm{Pe},\Phi_0) = \frac{\int_{r_{\mathrm{min}}}^{r_{\mathrm{max}}}\!\dif r\, r \int_{0}^{2\pi}\!\dif\phi_1 \int_{0}^{2\pi}\!\dif\phi_2 \, \abs{f_{\mathrm{app}}-f} }
{2\pi^{2} (r^{2}_{\mathrm{max}}-r^{2}_{\mathrm{min}}) } .
\end{equation}
In addition, we calculated the mean absolute value
\begin{equation}
\langle f \rangle(\mathrm{Pe},\Phi_0) = \frac{\int_{r_{\mathrm{min}}}^{r_{\mathrm{max}}}\!\dif r\, r \int_{0}^{2\pi}\!\dif\phi_1 \int_{0}^{2\pi}\!\dif\phi_2 \abs{f}} { 2\pi^{2} (r^{2}_{\mathrm{max}}-r^{2}_{\mathrm{min}}) } .
\end{equation}
A measure for the relative error of the analytic representation of $f$ is then given by the ratio $\mathrm{MAE}(\mathrm{Pe},\Phi_0)/\langle f\rangle(\mathrm{Pe},\Phi_0)$. The results for $\mathrm{MAE}(\mathrm{Pe},\Phi_0)$, $\langle f\rangle(\mathrm{Pe},\Phi_0)$, and $\mathrm{MAE}(\mathrm{Pe},\Phi_0)/\langle f\rangle(\mathrm{Pe},\Phi_0)$ are shown in Fig.\ \ref{fig:fit_error}. One can see that the relative error increases for low packing densities and either high or very low Pe, but never goes above $0.54$. The largest relative error occurs for $\mathrm{Pe}=10$ and $\Phi_0=0.04$. We found that the magnitude of the relative error is largely caused by the Fourier approximation \eqref{eq:Fourier_approximation}. Considering only the Fourier approximation, the relative error increases with $\mathrm{Pe}$ and reaches a maximum of $0.47$ at $\mathrm{Pe}=250$ and $\Phi_0=0.04$. The increase with $\mathrm{Pe}$ explains the similar behavior of the total relative error and results from the structure of the pair-distribution function becoming sharper when $\mathrm{Pe}$ grows. This sharpening originates from the weakening of thermal fluctuations for growing $\mathrm{Pe}$ and enlarges the contribution of higher-order Fourier modes, which are neglected in the approximation \eqref{eq:Fourier_approximation}. The increase of the relative error towards the origin at $\mathrm{Pe}=\Phi_0=0$ can also be found in the Fourier approximation and is amplified by the additional approximations. In particular, the skew $\lambda_I$ in the fit functions \eqref{eq:fit} for the Fourier coefficients becomes relatively small near the origin and starts to act predominantly as a second scaling parameter, which causes a conflict with the proper scaling parameter $f_{0,I}$. Moreover, the last fitting procedure introduces terms with divergences at $\mathrm{Pe}=0$ that amplify numerical errors for low \Peclet{} numbers. Furthermore, for very high and very low densities we observed an increasing statistical noise in some Fourier coefficients, which had a detrimental effect on the fitting procedures. The noise at very high densities correlates with the emergence of a hexatic phase that can be observed for sufficiently low \Peclet{} numbers\cite{DigregorioLSCGP2018} and breaks our initial assumption of isotropy in the approximation of the pair-distribution function.

\section{\label{sec:conclusions}Conclusions}
Based on Brownian dynamics simulations, we have studied the pair-distribution function of homogeneous suspensions of spherical ABPs in two spatial dimensions that interact through a WCA potential. We considered the full pair-distribution function with its dependence on a radial coordinate, two angular coordinates, the activity of the particles, and their overall packing density. An exploration of the properties of the pair-distribution function revealed that its structure can be explained by basic geometric and kinetic considerations. Furthermore, the general structure was found to be similar to that for hard-sphere ABPs reported in Ref.\ \onlinecite{HaertelRS2018}.    
We used the observed properties of the pair-distribution function to construct an approximate analytic expression for the product of the pair-distribution function and the interparticle force. This expression was found to be in good agreement with the simulation results.

The results for the pair-distribution function are helpful for the further theoretical investigation of systems of ABPs as well as nonequilibrium statistical physics in general. This is due to the fundamental importance of the pair-distribution function in the description of interactions in many-particle systems and for the development of field-theoretical models describing the collective dynamics of such systems \cite{WittkowskiSC2017, BickmannW2019, BialkeLS2013, ReinkenKBH2018, SpeckMBL2015, WittkowskiL2011,Bickmann2019a,BickmannBJW} We anticipate that the approximate analytic expression will lead to new advanced field theories for systems of active matter that go beyond those proposed in Refs.\ \onlinecite{BialkeLS2013, WittkowskiTSAMC2014, TiribocchiWMC2015, SpeckMBL2015, WittkowskiSC2017}. A first step in this direction has already been taken recently \cite{BickmannW2019}, where our analytic representation was used to calculate values of the coefficients occurring in a predictive field theory for interacting ABPs derived via the interaction-expansion method. The resulting predictions for, e.g., the spinodal demonstrate a significant gain in accuracy over previous results. Further work in this direction is currently in progress \cite{Bickmann2019a,BickmannBJW}.
Moreover, the analytic expression can be used as a reference case when developing analytic methods for predicting the pair-distribution function in active and other far-from-equilibrium systems. 
In the future, the procedure and methods used in this work can also serve as a template for investigations of other active systems, such as systems with different particle interactions, three spatial dimensions, other particle shapes, and mixtures between different types of particles.

\section*{Supplementary Material}
See Supplementary Material for a spreadsheet file containing the data for the characteristic length shown in Fig.\ \ref{fig:phase_diagram}, two movies showing the time evolution of a system of ABPs corresponding to a \Peclet{} number and packing density where the system remains homogeneous or forms clusters, respectively, a further movie corresponding to Fig.\ \ref{fig:corr_slices}a that shows the pair-distribution function as a function of its angular arguments and its evolution when the radial argument increases over time, a spreadsheet file containing the tables from the Appendix with the values of all fit parameters that are involved in the approximate analytic representation of the product function, as well as a Python script that imports the values of the fit parameters and provides a function for the analytic approximation of the product function.

\begin{acknowledgments}
R.W.\ is funded by the Deutsche Forschungsgemeinschaft (DFG, German Research Foundation) -- WI 4170/3-1. 
The simulations for this work were performed on the computer clusters PALMA and PALMA II of the University of M\"unster.
\end{acknowledgments}

\nocite{apsrev41Control}
\bibliographystyle{apsrev4-1}
\bibliography{control,refs}

\FloatBarrier
\onecolumngrid
\appendix
\section{Fit parameters for the analytic approximation of the function $\boldsymbol{-gU_{2}'}$}
In the following tables, the values of all fit parameters that are involved in the approximate analytic representation of the function $-g(r,\phi_1,\phi_2)U_{2}'(r)$ are given. The appropriate fit functions for the Fourier coefficients $f^{11}_{kl}$ and $f^{22}_{kl}$ follow from the corresponding parameter sets. For an easier use of the data, they are also available as a supplementary spreadsheet file. In addition, the Supplementary Material for this article contains a Python script that imports the parameter values from the spreadsheet file and provides a function for the analytic approximation of $-g(r,\phi_1,\phi_2)U_{2}'(r)$.

\setlength{\extrarowheight}{1.25pt}

\begin{table*}[htbp]
\centering
\begin{tabularx}{\linewidth}{|c|c|c|Y|Y|Y|Y|Y|Y|Y|}
\cline{4-10}
\csname @@input\endcsname fit_parameters_1
\end{tabularx}
\caption{\label{tab:fitparams_a0}Fit coefficients for the approximation of the Fourier coefficients $f^{11}_{00}$, $f^{11}_{01}$, and $f^{11}_{02}$ involved in the approximate analytic expression for the function $-g(r,\phi_1,\phi_2)U_{2}'(r)$.}
\end{table*}

\begin{table*}[htbp]
\centering
\begin{tabularx}{\linewidth}{|c|c|c|Y|Y|Y|Y|Y|Y|Y|}
\cline{4-10}
\csname @@input\endcsname fit_parameters_2
\end{tabularx}
\caption{\label{tab:fitparams_a1}Analogous to Tab.\ \ref{tab:fitparams_a0}, but now for $f^{11}_{10}$, $f^{11}_{11}$, and $f^{11}_{12}$.}
\end{table*}

\begin{table*}[htbp]
\centering
\begin{tabularx}{\linewidth}{|c|c|c|Y|Y|Y|Y|Y|Y|Y|}
\cline{4-10}
\csname @@input\endcsname fit_parameters_3
\end{tabularx}
\caption{\label{tab:fitparams_a2}Analogous to Tab.\ \ref{tab:fitparams_a0}, but now for $f^{11}_{20}$, $f^{11}_{21}$, and $f^{11}_{22}$.}
\end{table*}

\begin{table*}[htbp]
\centering
\begin{tabularx}{\linewidth}{|c|c|c|Y|Y|Y|Y|Y|Y|Y|}
\cline{4-10}
\csname @@input\endcsname fit_parameters_4
\end{tabularx}
\caption{\label{tab:fitparams_b1}Analogous to Tab.\ \ref{tab:fitparams_a0}, but now for $f^{22}_{11}$ and $f^{22}_{12}$.}
\end{table*}

\begin{table*}[htbp]
\centering
\begin{tabularx}{\linewidth}{|c|c|c|Y|Y|Y|Y|Y|Y|Y|}
\cline{4-10}
\csname @@input\endcsname fit_parameters_5
\end{tabularx}
\caption{\label{tab:fitparams_b2}Analogous to Tab.\ \ref{tab:fitparams_a0}, but now for $f^{22}_{21}$ and $f^{22}_{22}$.}
\end{table*}

\end{document}